\documentclass{article}

\usepackage[american]{babel}
\usepackage[utf8]{inputenc}
\usepackage{amsmath,amssymb,amsthm}
\usepackage{latexsym}
\usepackage{fullpage}
\usepackage[ruled,linesnumbered]{algorithm2e}
\newcommand{\madmx}{\textsc{madmx}}
\newcommand{\varun}{\textsc{varun}}
\newcommand{\rmasker}{\textsc{repeatmasker}}
\newcommand{\dna}{\textsc{dna}}
\newcommand{\HGO}{\textsc{hgmr}~1}
\newcommand{\HGF}{\textsc{hgmr}~5}
\newcommand{\sinealu}{\textsc{sine}/\textsc{alu}}
\newcommand{\lineone}{\textsc{line}/\textsc{l1}}

\newcommand{\BO}[1]{{O}\left(#1\right)}

\newtheorem{proposition}{Proposition}
\newtheorem{lemma}{Lemma}
\newtheorem{theorem}{Theorem}

\def\ME{{\cal M}}

\theoremstyle{definition}
\newtheorem{definition}{Definition}

\begin{document}

\title{\Large \bf MADMX: A Novel Strategy for Maximal Dense Motif Extraction\thanks{A preliminary version of this work was presented in WABI 2009.}}
\author{
Roberto Grossi\thanks{Dipartimento di Informatica,
Universit\`a di Pisa, Pisa, Italy. Email: \texttt{grossi@di.unipi.it}.}
\and 
Andrea Pietracaprina\thanks{Dipartimento di Ingegneria dell'Informazione,
Universit\`a di Padova, Padova, Italy. Email: \texttt{capri@dei.unipd.it}. Supported in part by the European Union under the FP6-IST/IP Project AEOLUS.}
\and 
Nadia Pisanti\thanks{Dipartimento di Informatica,
Universit\`a di Pisa, Pisa, Italy. Email:\texttt{pisanti@di.unipi.it}.}
\and 
Geppino Pucci\thanks{Dipartimento di Ingegneria dell'Informazione, Universit\`a di Padova, Padova, Italy. Email: \texttt{geppo@dei.unipd.it}. Supported in part by the European Union under the FP6-IST/IP Project AEOLUS.}
\and
Eli Upfal\thanks{Department of Computer Science, Brown University, Providence RI, USA.Email: \texttt{eli@cs.brown.edu}. Supported in part by the European Union under the FP6-IST/IP Project AEOLUS, NSF awards IIS-0325838 and DMI-0600384, and ONR Award N000140610607.}
\and
Fabio Vandin\thanks{Dipartimento di Ingegneria dell'Informazione, Universit\`a di Padova, Padova, Italy. Email: \texttt{vandinfa@dei.unipd.it} Contact Author. Supported in part by the European Union under the FP6-IST/IP Project AEOLUS. This work was done, in part, while the author was visiting the Department of Computer Science of Brown University.}
}

\maketitle 

\begin{abstract}

We develop, analyze and experiment with a new tool, called \madmx,
which extracts frequent motifs, possibly including don't care
characters, from biological sequences. We introduce \emph{density}, a
simple and flexible measure for bounding the number of don't cares in
a motif, defined as the ratio of solid (i.e., different from don't
care) characters to the total length of the motif.  By extracting only
\emph{maximal dense motifs}, \madmx\ reduces the output size and
improves performance, while enhancing the quality of the
discoveries. The efficiency of our approach relies on a newly defined
combining operation, dubbed \emph{fusion}, which allows for the
construction of maximal dense motifs in a bottom-up fashion, while
avoiding the generation of nonmaximal ones. We provide experimental
evidence of the efficiency and the quality of the motifs returned by
\madmx.
\end{abstract}

\section{Introduction}
\label{sec:introduction}

The discovery of frequent patterns (\emph{motifs}) in biological
sequences has attracted wide interest in recent years, due to the
understanding that sequence similarity is often a necessary condition
for functional correlation. Among other applications, motif discovery
proves an important tool for identifying regulatory regions and
binding sites in the study of functional genomics. From a
computational point of view, a major complication for the discovery of
motifs is that they may feature some sequence variation without loss
of function. The discovery process must therefore target
\emph{approximate motifs}, whose occurrences are similar but not
necessarily identical.  Approximate motifs are often modeled through
the use of the \emph{don't care} character in certain positions, which
is a wild card matching all characters of the alphabet, called \emph{solid
  characters} \cite{Parida08}.

Finding interesting approximate motifs is computationally
challenging. As the number of don't cares increases
and/or the minimum frequency threshold decreases, the output may
explode combinatorially, even if the discovery targets only maximal
motifs---a subset of the motifs which implicitly represents the
complete set. Moreover, even when the final output is not too large,
partial data during the inference of target motifs might lead to
memory saturation or to extensive computation during the intermediate
steps.

\sloppy
A large body of literature in the last decade has dealt with
efficient motif discovery
\cite{Parida00,ApostolicoP04,PisantiCGS05,ApostolicoT07a,Ukkonen07,MorrisNU08,ArimuraU08,ApostolicoT08,ApostolicoCP09},
and an excellent survey of known results can be found in the book
\cite{Parida08}. In order to alleviate the computational burden of
motif extraction and to limit the output to the most promising or interesting discoveries, some works combine the traditional use of a
frequency threshold with restrictions on the flexibility of the
extracted motifs, often captured by limitations on the number of
occurring don't cares. 

In a recent work, Apostolico et al.~\cite{ApostolicoCP09} study the
extraction of \emph{extensible motifs}, comprising standard don't
cares and extensible wild cards. The latter are spacers of variable
length that can take different size (within pre-specified limits) in
each occurrence of the motif.  An efficient tool, called \varun, is
devised in \cite{ApostolicoCP09} for extracting all maximal extensible
motifs (according to a suitable notion of maximality defined in the
paper) which occur with frequency above a given threshold $\sigma$ and
with upper limits $D$ on the length of the spacers.  \varun\ returns
the extracted motifs sorted by decreasing z-score, a widely adopted
statistical measure of interestingness. The authors demonstrate the
effectiveness of their approach both theoretically, by proving that
each maximal motif features the highest z-score within the class of
motifs it represents, and experimentally, by showing that 
the returned top-scored motifs comprise biologically relevant ones
when run on protein families and \dna\ sequences.
 
A slightly more general way of limiting the number of don't cares in a
motif has been explored in \cite{RigoutsosF98}. The authors define $\langle L,W
\rangle$ motifs, for $L \leq W$, where at least $L$ solid characters
must occur in each substring of length $W$ of the motif. They propose a
strategy for extracting $\langle L,W \rangle$ motifs which are also
maximal, although their notion of maximality is not internal to the
class of $\langle L,W \rangle$ motifs. As a consequence, the algorithm
is not complete, since it disregards all those $\langle L,W \rangle$
motifs that are subsumed by a maximal non-$\langle L,W \rangle$ one.

\paragraph*{\bf Our results.} 
Our work focuses on the discovery of \emph{rigid motifs}, which
contain blocks of solid characters (solid blocks) separated by one
or more don't cares.  We propose a more general approach for
controlling the number of don't cares in rigid motifs.  Specifically,
we introduce the notion of \emph{dense motif}, a frequent pattern
where the fraction of solid characters is above a given threshold. Our
density notion is more flexible and general than the one considered in
\cite{Parida08,ApostolicoCP09}, since it allows for arbitrarily long
runs of don't cares as long as the fraction of solid characters in the
pattern is above the threshold. We define a natural notion of
\emph{maximality} for dense patterns and devise an efficient
algorithm, called \madmx\ (pronounced \emph{Mad Max}), which performs
complete \textsc{ma}ximal \textsc{d}ense \textsc{m}otif
e\textsc{x}traction from an input sequence, with respect to
user-specified frequency and density thresholds.

The key technical result at the core of our extraction strategy is a
closure property which affords the complete generation of all maximal
dense motifs in a breadth-first fashion, through an
\emph{apriori}-like strategy \cite{AgrawalS94-1}, starting from a
relatively small set of solid blocks, and then repeatedly applying a
suitable combining operator, called \emph{fusion}, to pairs of
previously generated motifs.  In this fashion, our strategy avoids the
generation and consequent storage of intermediate patterns which are
not in the output set, which ensures time and space complexities
polynomial in the combined size of the input and the output.

We performed a number of experiments on \madmx\ to assess the
biological significance of maximal dense motifs and to compare \madmx\
against its most recent and close competitor \varun.  For the first objective, we
used \madmx\ to extract maximal dense motifs from a number of human
\dna\ fragments. We compared the output set against those in
\texttt{RepBase} \cite{repbase}, the largest repository of repetitive
patterns for eukaryotic species, using \rmasker\ \cite{repeatmasker},
a popular tool for masking repetitive \dna. The experiments show that
all of our returned motifs are occurrences of patterns in
\texttt{RepBase}, and \emph{fully} characterize the family of \sinealu\
repeats (and partially the \lineone\ family). This provides evidence
that the notion of density, when applied to rigid motifs, captures
biological significance.

Next we compared the z-score performance of \madmx\ and \varun. We ran
both algorithms on several families of \dna\ fragments, limiting
\varun\ to the generation of rigid motifs and setting the parameters
so as to obtain comparable output sizes, with motifs listed by
decreasing z-score. The experiments show that the top-$m$
highest-ranking motifs returned by \madmx\ almost always feature
higher z-scores than the corresponding top-$m$ ones returned by \varun,
even for large values of $m$, with only a modest increase in running
time, which may be partly due to the fact that coding of \madmx\ is yet
to be optimized.  In fairness, we must remark that \varun\ deals also
with extensible motifs while \madmx\ only targets rigid motifs.

The paper is organized as follows. In Section~\ref{sec:prelim} several
technical definitions and properties of motifs with don't cares are
given. Section~\ref{sec:maximal} proves the closure property at the
base of \madmx\ and provides a high-level description of the
algorithm.  In Section~\ref{sec:exp}, the experimental validation of
\madmx\ is presented.

\section{Preliminary Definitions and Properties}
\label{sec:prelim}

Let $\Sigma$ be an alphabet of $m$ characters and let $s = s[0] s[1]
\dots s[n-1] $ be a string of length $n$ over $\Sigma$. We use $s[i
\dots j ]$ to denote the substring $s[i]$ $s[i+1]$ $\cdots$ $s[j]$ of
$s$, for $i\leq j$. Characters in $\Sigma$ are also called \emph{solid
  characters}. We use $\circ \not\in \Sigma$ to denote a distinguished
character called \emph{wild card} or \emph{don't care} character.  Let
$\epsilon$ denote the empty string. A \emph{pattern} $x$ is a string
in $\{\epsilon\} \cup \Sigma \cup \Sigma(\Sigma \cup \{ \circ \})^*
\Sigma$. However, whenever necessary, we will assume that patterns are 
implicitly padded to their left and right with arbitrary sequences of don't care
characters. 

Given two patterns $x, y$ we say that $y$ is \emph{more
  specific} than $x$, and write $x \preceq y$, iff for every $i \geq 0$
either $x[i]=y[i]$ or $x[i] = \circ$. 
Given two patterns $x, y$ we say that $x$ \emph{occurs in $y$ at
  position $\ell$} iff $x \preceq y[\ell \ldots \ell+|x|-1]$: we also
say that $y$ \emph{contains}~$x$.
For a string $s$, the \emph{location list} ${\cal L}_x$ of a pattern
$x$ in $s$ is the complete set of positions at which $x$ occurs in
$s$.  We refer to $f(x) = |{\cal L}_x|$ as the \emph{frequency} of
pattern $x$ in $s$. (Note that $f(\epsilon)=n$.)  As in
\cite{Ukkonen07}, the \emph{translated representation} of the location
list ${\cal L}_x=\{l_0,l_1,l_2,\dots,l_k\}$ is $\tau{(\cal
  L}_x)=\{l_1-l_0,l_2-l_0,\dots,l_k-l_0\}$.  Given two patterns $x,y$,
we say that $y$ \emph{subsumes} $x$ in $s$ if $f(x)=f(y)$ and $y$
contains $x$. As a consequence, if $y$ subsumes $x$ then $\tau({\cal
  L}_x)=\tau({\cal L}_y)$. A pattern $x$ is \emph{maximal} if it is
not subsumed by any other pattern $y$. (We observe that this notion of
maximality coincides with that of \cite{PisantiCGS05}.)  Given a
pattern $x$, its \emph{maximal extension} $\ME(x)$ is the maximal
pattern that subsumes $x$, which can be shown to be unique
\cite{PisantiCGS05}.

In what follows, we call \emph{solid block} a string in $\Sigma^{+}$
and a \emph{don't care block} a string in $\circ^{+}$. Furthermore,
given a pattern $x$, $\mbox{dc}(x)$ denotes the number of don't care
characters contained in $x$. 

\begin{definition}
The \emph{density} $\delta(x)$ of $x$ is:
$\delta(x)=1-\mbox{dc}(x)/|x|$.  Given a (density) threshold $\rho$,
$0 <\rho \leq 1$, we say that a pattern $x$ is \textit{dense} if
$\delta(x) \geq \rho$.
\end{definition}
Note that a solid block is a dense pattern with respect to every
threshold $\rho$.  

It is reasonable to concentrate the attention on dense patterns that
are not subsumed by any other dense pattern, since they are the most interesting dense representatives 
in the equivalence classes induced by ``sharing'' the same translated representation; these representatives are defined below.

\begin{definition}\label{def:maxdense}
A dense pattern $x$ is a \emph{maximal dense pattern} in $s$ if it is
not subsumed by any other dense pattern $x' \neq x$.
\end{definition}
Observe that a maximal dense pattern $x$ needs not be a maximal
pattern in the general sense, since $\ME(x)$ might be a nondense
pattern. However, every dense pattern $x$ is subsumed by \emph{at
  least} one maximal dense pattern.  In fact, all of the maximal dense
patterns that subsume $x$ are dense substrings of $\ME(x)$, namely,
those that contain $x$ and are not substrings of any other dense
substring of $\ME(x)$. We want to stress that there might be several
maximal dense patterns that subsume~$x$.  As an example, for
$\rho=2/3$, the dense pattern $x=\mathtt{B}$ in the string
$S=\mathtt{AdBeCfAgBhC}$ is subsumed by maximal dense patterns
$\mathtt{A} \circ \mathtt{B}$ and $\mathtt{B} \circ \mathtt{C}$, while
$\ME(x) = \mathtt{A} \circ \mathtt{B} \circ \mathtt{C}$ is not dense.

\begin{definition}
Given a frequency threshold $\sigma$ and a density threshold $\rho$, a
pattern $x$ is a \emph{dense maximal motif} in $s$ if $x$ is a maximal
dense pattern in $s$ with respect to $\rho$, and $f(x)\geq \sigma$.
A dense maximal motif for $\rho=1$ is also referred to as
\emph{maximal solid block}. 
\end{definition}

\noindent
\textbf{Problem of interest.}
We are given an input string $s$, a frequency threshold $\sigma$, 
and a density threshold $\rho$. Find all the maximal dense motifs in $s$.

\smallskip

In the rest of the paper, we will omit referencing
the input string $s$ when clear from the context. An important
property of maximal dense patterns, which we will exploit in our
mining strategy, is that all of their solid blocks are maximal solid
blocks. This property is stated in the following proposition whose
proof, omitted for brevity, extends a similar result holding for
arbitrary maximal patterns \cite{Ukkonen07,Pisanti02}.
\begin{proposition}\label{maximalpatterns}
Let $x$ be a maximal dense pattern with respect to a density threshold
$\rho$, and let $b=x[i \ldots j]$  be a solid block
in $x$ such that $x[i-1]=x[j+1]=\circ$ and $j \geq i$. Then, $b$ is a maximal solid block.
\end{proposition}

\section{An Algorithm for MAximal Dense Motif eXtraction}
\label{sec:maximal}

In this section we describe our algorithm, called \madmx\
(pronounced \emph{Mad Max}), for \textsc{ma}ximal \textsc{d}ense
\textsc{m}otif e\textsc{x}traction. The algorithm adopts a
breadth-first 
  \emph{apriori}-like strategy \cite{AgrawalS94-1}, similar in spirit to the
  one developed in \cite{ApostolicoCP09}, using maximal solid blocks
  as building blocks by Proposition~\ref{maximalpatterns}. \madmx\ operates by repeatedly
  combining together, in a suitable fashion, pairs of maximal dense
  motifs, and extracting from the combinations
less frequent maximal dense motifs.

A key notion for the algorithm, underlying the aforementioned
combining operations, is the \textit{fusion} of characters/patterns. 
\begin{definition}
Given three characters $c,c_1,c_2 \in \Sigma \cup \{\circ\}$, we say that $c$ is the fusion of $c_1$ and $c_2$, and write $c = c_1 \bigtriangledown c_2$,
if one of the following holds:
\begin{enumerate}
 \item $c = c_1 = c_2$;
 \item $c_1 = \circ$, $c = c_2 \neq \circ$;
 \item $c = c_1 \neq \circ$, $c_2 = \circ$.
\end{enumerate}
\end{definition}
The above notion of fusion generalizes to patterns as follows.
\begin{definition}
Given three patterns $x,y,z$ and an integer $d$, we say that $z$ is
the \emph{$d$-fusion of $x$ and $y$}, and write $z=x
\bigtriangledown_d y$, if $z$ can be obtained by removing the leading and
trailing don't care characters from the pattern $m$ defined as $m[i]
= x[i+d] \bigtriangledown y[i]$, for all indices $i$.
\end{definition}

The breadth-first strategy adopted by our algorithm crucially 
relies on the following theorem, which highlights the structure of dense
motifs:

\begin{theorem}\label{theorem:completeness}
Let $x$ be a maximal dense motif with $\mbox{dc}(x)>0$. Then:
\begin{enumerate}
\item[(a)] there exists a maximal solid block $b$ in $x$ 
such that $\ME(x)=\ME(b)$, or
\item[(b)] there exist two maximal dense motifs $y_1,y_2$ such that:
\begin{itemize}
\item $\ME(x)=\ME(y_1 \bigtriangledown_d y_2)$, for some $d$;
\item there are two maximal solid blocks $b_1,b_2$ in $x$ and an
  integer $\hat{d}>0$ such that $b_1$ is a maximal solid block in
  $y_1$, $b_2$ is a maximal solid block in $y_2$, and $b_1
  \circ^{\hat{d}} b_2$ is contained in $y_1\bigtriangledown_dy_2$;
\item $f(x) < \min\{f(y_1),f(y_2)\}$;
\end{itemize}
\end{enumerate}
\end{theorem}
For the proof of Theorem~\ref{theorem:completeness} we need to define
another type of pattern combination, namely the operation of
\textit{merge} between two patterns, which is similar to the one
introduced in \cite{PisantiCGS05}. Given two characters $c_1,c_2$,
we define the operator $\oplus$ between them such
that $c_1 \oplus c_2 = \circ$, if $c_1 \neq c_2$, and 
$c_1 \oplus c_2 = c_1 = c_2$, otherwise.
\begin{definition}
Given two patterns $x,y$ and an integer $d$, the $d$-merge of $x$ and
$y$ is the pattern $z = x \oplus_d y$ which can be obtained by removing
all leading and trailing don't cares from the pattern $m$
defined as $m[i] = x[i+d] \oplus y[i]$ for all $i$.
\end{definition}
We want to stress the difference between the notions of merging and
fusion: the merge of two patterns $x,y$ is always well defined and more 
general than $x,y$, while the fusion of $x,y$ may not exist and, if it does,
is more specific than $x,y$.

For the proof of Theorem~\ref{theorem:completeness} we also need the
property established by the following lemma.
\begin{lemma}\label{lem:maxmax}
Let $x$ and $y$ be maximal patterns, and $d$ be an integer such that $z = x
\oplus_d y \neq \epsilon$. Then $z$ is a maximal pattern. Moreover, if
$z \neq x$ (resp., $z \neq y$) then $f(z) > f(x)$ (resp., $f(z) >
f(y)$).
\end{lemma}
\begin{proof}
First we prove that $z$ is maximal.  By contradiction, suppose that
this is not the case. Then, there exists a position $i$ such that
$z[i]=\circ$ and we can replace the $\circ$ with a solid character $c$
without decreasing the frequency of the pattern. (Note that the
position of the substitution can be to the left of the first character
in $z$ or to the right of the last character in $z$.) Since $x$ and
$y$ are more specific than $z$, to every occurrence of $x$ and $y$ in
the string corresponds an occurrence of $z$. Hence, every occurrence of
$x$ (resp., $y$) in the string, contains $c$ in its $i+d$th (resp.,
$i$th) position. Therefore, by maximality of $x$ and $y$, it must be
$z[i]=x[i+d]=y[i]=c$, which is a contradiction. The relations between
the frequencies of $x, y$ and $z$ follow trivially by their
maximality. \qed
\end{proof}
We are now ready to prove the theorem. 

\begin{proof}[Theorem~\ref{theorem:completeness}]
Given a pattern $x$ and two nonnegative integers $i \leq j$, we let
$x^*[i\dots j]$ denote the pattern obtained by removing all the
leading and trailing don't care characters from $x[i\dots j]$.  Since
$x$ is a maximal dense pattern and $dc(x)>0$, it is easy to see that
there exist two dense patterns $x_1,x_2$ and an integer $d>0$ such
that $x = x_1 \circ^d x_2$, hence there exists an index
$s_1>0$ such that $x^*[0 \dots s_1 - 1]$ and $x^*[s_1+1 \dots |x|-1]$
are dense. We call these two patterns the \textit{level-1
  decomposition} of $x$ (observe that many such decompositions may
exist). Also, we let $\ell_1 = 0$ and $r_1 =
|x|-1$. Now, consider the following iterative process:
\begin{enumerate}

\item\label{step1} If in the level-$i$ decomposition of $x$ both
  $x^*[\ell_i\dots s_i-1]$ and $x^*[s_i+1\dots r_i]$ have frequency
  strictly greater than $f(x)$, \textit{or} at least one of
  $x^*[\ell_i\dots s_i-1]$ and $x^*[s_i+1\dots r_i]$ is a solid block
  with frequency equal to $f(x)$, then terminate;
 \item Otherwise, let $y = x^*[\ell_{i+1}\dots r_{i+1}]$ be (an
   arbitrary) one of $x^*[\ell_i\dots s_i-1]$ or $x^*[s_i+1\dots r_i]$
   which is not a solid block and has frequency equal to $f(x)$. Since
   $y$ is dense, there exists an index $s_{i+1}$, $\ell_{i+1} <
   s_{i+1} < r_{i+1}$ such that $x^*[\ell_{i+1} \dots s_{i+1}-1]$ and
   $x^*[s_{i+1}+1 \dots r_{i+1}]$ are both dense. Call these two
   patterns the level-$(i+1)$ decomposition of $x$.  Set $i = i+1$ and  go to Step~\ref{step1}.
\end{enumerate}

Assume that the decomposition process ends by finding a solid block
$b$ that is a solid block in $x$ and has $f(b)=f(x)$. Then, $\ME(b) =
\ME(x)$ and the theorem follows. Otherwise, at the last level $j$ of
the decomposition, we have that $f(x) < \min\left\{f(x^*[\ell_j\dots
  s_j-1]), f(x^*[s_j+1\dots r_j])\right\} $. In this latter case, as
explained in Section~\ref{sec:prelim} (after
Definition~\ref{def:maxdense}), we can determine two maximal dense
patterns $y_1,y_2$ such that $y_1$ contains $x^*[\ell_j\dots s_j-1]$,
$y_2$ contains $x^*[s_j+1\dots r_j]$, and with $\ME(y_1) =
\ME(x^*[\ell_j\dots s_j-1])$ and $\ME(y_2) = \ME(x^*[s_j+1\dots
  r_j])$. Since $f(y_1) = f(x^*[\ell_j\dots s_j-1])$ and $f(y_2) =
f(x^*[s_j+1\dots r_j])$, we have that $f(x) <
\min\left\{f(y_1),f(y_2)\right\}$. Observe that by construction there
must exist two solid blocks $b_1,b_2$ in $x$ and an integer $\hat{d}$
such that $b_1$ is a solid block in $y_1$, $b_2$ is a solid block in
$y_2$, and $b_1 \circ^{\hat{d}} b_2$ is a sequence of two solid blocks
in $x$. In fact, $b_1$ (resp., $b_2$) is the last (resp., the first) solid
block of $x^*[\ell_j\dots s_j-1]$ (resp., $x^*[s_j+1\dots r_j]$).

Next, we show that there exists a $d$ such that the $d$-fusion $y_1
\bigtriangledown_d y_2$ is well defined, contains $b_1 \circ^{\hat{d}}
b_2$, and $\ME(y_1 \bigtriangledown_d y_2) = \ME(x)$. We proceed as
follows.  Let us ``align'' $\ME(x)$ and $y_1$ so to match the occurrences
of $b_1$ in both patterns. Then, for a certain integer
$p$, $\ME(x)[i+p]$ corresponds to $y_1[i]$. Assume, for the sake
of contradiction, that there exists an index $j$ such that
$\ME(x)[j+p]$ is not more specific than $y_1[j]$.  Then,
Lemma~\ref{lem:maxmax} implies that $z = \ME(x) \oplus_p \ME(y_1) \neq
\ME(y_1)$, which contains $x^*[\ell_j\dots s_j-1]$, is maximal and has
frequency strictly greater than $f(y_1)$, which is impossible because
we have chosen $y_1$ such that $\ME(x^*[\ell_j\dots s_j-1]) =
\ME(y_1)$ and therefore $f(x^*[\ell_j\dots s_j-1])= f(y_1)$. Therefore,
$\ME(x)$ contains $y_1$. A similar argument shows that
$\ME(x)$ contains $y_2$.

Since $y_1$ and $y_2$ are contained in $\ME(x)$, 
there must exist a $d$ such that $y_1 \bigtriangledown_d y_2$
is well defined and can be aligned with $\ME(x)$ in such a way to match
the blocks $b_1$ and $b_2$ of $y_1$ and $y_2$ with the corresponding
blocks in $\ME(x)$. Moreover, $\ME(x)$ contains
$y_1 \bigtriangledown_d y_2$, hence $f(y_1 \bigtriangledown_d y_2) 
\geq f(\ME(x)) = f(x)$. However, since $y_1 \bigtriangledown_d y_2$
contains both $x^*[\ell_j\dots s_j-1]$ and $x^*[s_j+1\dots r_j]$,
it contains also  $x^*[\ell_j\dots r_j]$, which, by the decomposition
process, has frequency equal to $f(x)$. Therefore, 
$f(y_1 \bigtriangledown_d y_2) \leq f(x)$, and the theorem follows since
$f(y_1 \bigtriangledown_d y_2) =f(x)$.  \qed
\end{proof}

In essence, Theorem~\ref{theorem:completeness} guarantees that we can
find any maximal dense motif $x$ either within $\ME(b)$, for some
maximal solid block $b$, or by $d$-fusing two higher-frequency maximal
dense motifs $y_1,y_2$, for some $d$, finding
$z=\ME(y_1\bigtriangledown_d y_2)$ and then possibly ``trimming'' $z$
on both sides to obtain $x$. 

\begin{figure}[t]
\begin{center}
\begin{algorithm}[H]
\begin{small}
\SetKw{Whilef}{while}
\SetKw{Dof}{do}
\SetKw{Thenf}{then}
\SetKw{Forf}{for}
\SetKw{Iff}{if}
\caption{\madmx \label{alg:maxdense}}
\KwIn{String $s$, frequency threshold $\sigma$, density threshold $\rho$} 
\KwOut{Maximal dense motifs} 
\BlankLine
$\mathit{previous} \leftarrow \emptyset$, $\mathit{current} \leftarrow \emptyset$, $\mathit{next} \leftarrow \emptyset$ \;
$\mathit{blocks}$ $\gets$ maximal solid blocks of $s$ with frequency $\geq \sigma$\; \label{codemax:init}
\Forf {\bf each} $b \in \mathit{blocks}$ \Dof \label{codemax:forinit}\\
	\Indp 
	find $\ME(b)$ \;
	\label{codemax:2} ${\cal DM}\leftarrow$ extractMaximalDense($\ME(b)$)\; 
	\Forf {\bf each} $x \in {\cal DM}$ \Dof {$\mathit{current} \leftarrow \mathit{current} \cup \{x\}$\; \label{codemax:forend}}
	\Indm
\Whilef $\mathit{current} \neq \emptyset$ \Dof \label{codemax:whileinit}\\
	\Indp 
	\Forf {\bf each} $x_1 \in \mathit{current}$ \Dof \label{codemax:1}\\
		\Indp 
		\Forf {\bf each} $x_2 \in \mathit{previous} \cup \mathit{current}$ \Dof\label{codemax:1a} \\
			\Indp 
				\Forf {\bf each} $d$ s.t. $z = x_1
                                \bigtriangledown_d x_2$ is a valid fusion \Dof \label{codemax:fusion}\\
 				\Indp 
					find $\ME(z)$\;
\label{codemax:3} ${\cal DM} \leftarrow$ extractMaximalDense($\ME(z)$)\;
					\Forf {\bf each} $x \in {\cal DM}$ \Dof\\
					\Indp
						\lIf{$f(x) \geq \sigma$ and $x \notin \mathit{previous} \cup \mathit{current} $}{$\mathit{next} \leftarrow \mathit{next} \cup \{x\}$\;}
					\Indm
				\Indm
			\Indm
		\Indm
		$\mathit{previous} \gets \mathit{previous} \cup \mathit{current}$\; 
		$\mathit{current} \gets \mathit{next}$; $\mathit{next}\gets\emptyset$\; \label{codemax:whileend}
	\Indm
\Return $\mathit{previous}$\;
\end{small}
\end{algorithm}
\end{center}
\caption{Pseudocode of algorithm \madmx.} \label{code:maxdense}
\end{figure}

\sloppy Algorithm~\madmx, whose pseudocode is reported in
Figure~\ref{code:maxdense}, implements the strategy inspired by
Theorem~\ref{theorem:completeness}. It employs three (initially empty)
sets \emph{previous}, \emph{current}, and \emph{next}. In
Line~\ref{codemax:init}, the algorithm first stores the maximal solid
blocks $b$ in $s$ for the given frequency in the set \emph{blocks}
(see Section~\ref{sec:prelim}). Then, it extracts all of the
appropriate maximal dense motifs from $\ME(b)$ in
Lines~\ref{codemax:forinit}--\ref{codemax:forend}, using the function
extractMaximalDense, as implied by
Theorem~\ref{theorem:completeness}(a).  Finally,
Lines~\ref{codemax:whileinit}--\ref{codemax:whileend} implement the
strategy as implied by Theorem~\ref{theorem:completeness}(b). 
(In Line~\ref{codemax:fusion} a $d$-fusion $y_1\bigtriangledown_d y_2$ is considered \emph{valid}
if it satifies the second property of Theorem~\ref{theorem:completeness}(b).)

An important issue for the efficiency of \madmx\ is that it needs to
compute the exact frequency of each generated pattern. For what
concerns the fusion operation of two patterns $x_1, x_2$ in
Line~\ref{codemax:fusion}, observe that a simple computation on the
pairs $(\ell_1,\ell_2)\in {\cal L}_{x_1} \times {\cal L}_{x_2}$ is
sufficient to yield the frequencies of all the valid fusions of two
patterns.  However, given $z =x_1\bigtriangledown_d x_2$, for a
maximal dense pattern $w$ which does not contain $z$ in its entirety,
we can only conclude that $f(w) \geq f(z)$. We then label the motifs
for which the exact frequencies are known as \emph{final}, and those
for which only a lower bound to their frequencies is known as
\emph{tentative}, and update the lower bounds and the labels during
the execution of the algorithm. Whenever the set \emph{current}
contains no final motifs, we can label as final the tentative motif in
\emph{current} with the highest lower bound to its frequency, and
continue with the generation.  The proof of the correctness of this
assumption and further details on the implementation of the algorithm
will be provided in the full version of this extended abstract.
A crude upper bound on the running time of \madmx\ can be derived by
observing that, for each pair of dense maximal motifs in output, the
time spent during all the operations concerning that pair is (naively)
$\BO{n^3}$, where $n$ is the length of the input string. If $P$
patterns are produced in output, the overall time complexity is
$\BO{n^3 P^2}$.

\section{Experimental Validation of MADMX}
\label{sec:exp}

We developed a first, non-optimized, implementation of \madmx\ in \texttt{C++}
also including an additional feature which eliminates, from the set of
initial maximal solid blocks, those shorter than a given threshold
$min_\ell$. The purpose of this latter heuristics is to speed up motif
generation driving it towards the discovery of (possibly) more
significant motifs, with the exclusion of spurious, low-complexity
ones. (The code is available for download at {\tt
  http://www.dei.unipd.it/wdyn/?IDsezione=4534}.)

We performed two classes of experiments to evaluate how significant is
the set of motifs found using our approach. The first class of
experiments, described in Section~\ref{sub:eval-repbase}, compares our
motifs with the known biological repetitions available in
\texttt{RepBase} \cite{repbase}, a very popular genomic database.  The
second class of experiments, described in
Section~\ref{sub:eval-z-score}, aims at comparing the motifs extracted
by \madmx\ with those extracted by \varun\ using the same $z$-score
metric employed in \cite{ApostolicoCP09} for assessing their relative
statistical significance.

\subsection{Evaluating significance by known biological repetitions}
\label{sub:eval-repbase}

\texttt{RepBase} \cite{repbase} is one of the largest repositories of
prototypic sequences representing repetitive \dna\ from different
eukaryotic species, collected in several different ways.
\texttt{RepBase} is used as a reference collection for masking and
annotation of repetitive \dna\ through popular tools such as
\rmasker\ \cite{repeatmasker}. \rmasker\ screens an input
\dna\ sequence $s$ for simple repeats and low complexity portions, and
interspersed repeats using \texttt{RepBase}. Sequence comparisons are
performed through Smith-Waterman scoring. \rmasker\ returns a detailed
annotation of the repeats occurring in $s$, and a modified version of
$s$ in which all of the annotated repeats are masked by a special
symbol (\texttt{N} or \texttt{X}).  With the current version of \texttt{RepBase}, on average,
almost 50\% of a human genomic \dna\ sequence will be masked by the
program \cite{repeatmasker}.

Most of the interspersed repeats found by \rmasker\ belong to the
families called \sinealu\ and \lineone: the former are \emph{Short
  INterspersed Elements} that are repetitive in the \dna\ of
eukaryotic genomes (the Alu family in the human genome); the latter
are \emph{Long Interspersed Nucleotide Elements}, which are typically
highly repeated sequences of 6K--8K bps, containing \textsc{rna}
polymerase II promoters. The \lineone\ family forms about 15\% of the
human genome.

We have conducted an experimental study using \madmx\ and \rmasker\ on
\emph{Human Glutamate Metabotropic Receptors} \HGO\ (410277 bps) and
\HGF\ (91243 bps) as input sequences. We have downloaded the
sequences from the March 2006 release of the UCSC Genome database
(\texttt{http://genome.ucsc.edu}).
\noindent \rmasker\ version was open-3.2.7, sensitive mode, with the
query species assumed to be homologous; it ran using
\texttt{blastp} version 2.0a19MP-WashU, and \texttt{RepBase} update
20090120.

The experiments to assess the biological significance of the maximal
dense motifs extracted by \madmx\ involved three separate stages.  In
the first stage, we ran \rmasker\ on the input sequences \HGO\ and
\HGF, searching for interspersed repeats using \texttt{RepBase}. One
of the output files (\texttt{.out}) of \rmasker\ contains the list of found repeats,
and provides, for each occurrence, the substring $s[i \ldots j]$ of
the input sequence $s$ which is locally aligned with (a substring of) the
repeat.

In the second stage, we ran \madmx\ on the same DNA sequences, with
density threshold $\rho=0.8$, frequency threshold $\sigma = 4$, and
$\min_\ell = 15$.  In order to filter out simple repeats and low
complexity portions, which are dealt with by \rmasker\ without
resorting to \texttt{RepBase}, we modified \madmx\ eliminating
periodic maximal solid blocks (with short periods), which are the
seeds of simple repeats. Then, we identified the occurrences of the
motifs returned by \madmx\ in the input sequences, using \rmasker\ as
a pattern matching tool (i.e., replacing \texttt{RepBase} with the set
of motifs returned by \madmx\ as the database of known repeats). The
underlying idea behind this use of \rmasker\ was to employ the same
local alignment algorithms, so to make the comparison fairer.

In the third stage, we cross-checked the intervals associated with the
occurrences of the \texttt{RepBase} repeats against those associated
with the occurrences of our motifs. Surprisingly, \madmx\ was able to
identify and characterize \emph{all} of the intervals of the known
\sinealu\ repeats in \HGO\ and \HGF\ (respectively, 56 repeats plus an
extra unclassified for \HGO, and 20 plus an extra unclassified for
\HGF). The remaining occurrences of the motifs permitted to identify
29 repeats out of 78 of the \lineone\ family in \HGO. (A more detailed
account of the whole range of experiments conducted using
\rmasker\ and the data sets by Tompa et el.\mbox{} and Sandve et al.\mbox{} 
will be provided in the full version.)

\subsection{Evaluating significance by statistical z-score ranking}
\label{sub:eval-z-score}

The z-score is the measure of the distance in standard deviations of
the outcome of a random variable from its expectation.  Consider a
\dna\ sequence $s$ of length $n$ as if it was generated by a
stationary, i.i.d. source with equiprobable symbols; an
approximation to the z-score for a
motif of length $m$ that contains $c$ solid characters and appears $f$
times in $s$ is given by $Z = \frac{f - (n-m+1) \times
  p}{\sqrt{(n-m+1) \times p \times (1-p)\,}}$, where $p=(1/4)^c$.
This metric was used in \cite{ApostolicoCP09} to assess the
significance of the motifs extracted by \varun\ and to rank them in
the output.

We employed the code for \varun\ provided by the authors to extract
the rigid motifs from the \dna\ sequences analyzed in
\cite{ApostolicoCP09}.  We then ran \madmx\ on the same sequences
using the same frequency parameters, and setting the minimum density
threshold $\rho$ in such a way to obtain a comparable yet smaller
output size. In this fashion, we tested the ability of \madmx\ to
produce a succinct yet significant set of motifs, by virtue of its
more flexible notion of density.

The results are shown in Table~\ref{tab:madmx&varun}. For \varun\ we
used $D=1$, thus allowing at most one don't care between two solid
characters, and ran \madmx\ with $min_\ell=1$, so to obtain the
\emph{complete} family of maximal dense motifs.  In the table, there
is a row of the table for each sequence (identified in the first
column).  Each sequence, whose total length is reported in the second
column, is obtained as the concatenation of a number of smaller
subsequences, reported in the third column. On each sequence, both
tools were run with the same frequency threshold $\sigma$, and the
table reports for both the output size in terms of the number of
motifs returned and the execution time in seconds. Also, for \madmx, the table
reports the density threshold $\rho$ used in each experiment.

\begin{table}[h]
\begin{center}
\begin{tabular}{|l|c|c|c||c|c||c|c|c||c|c|c|c|c|}
\hline
& &  & & \multicolumn{2}{|c||}{{\varun}} &
\multicolumn{3}{|c||}{{\madmx}} & \multicolumn{5}{|c|}{best top-$m$ z-scores} \\
\cline{5-14}
name & length & \# & $\sigma$ & \,$|$output$|$\, & time & $\rho$ & \,$|$output$|$\, & time & $m$=10 & $m$=50 & $m$=100 & $m^*$ & $\hat{m}$\\
\hline
\hline
\texttt{ace2} & 500 & 1 & 2 & 1866 & 3s & 0.7 & 1762 & 18s & 10 & 50 & 100 & 1571 & 1067\\
\hline
\texttt{ap1} & 500 & 1 & 2 & 1555 & 1s & 0.7 & 1304 & 5s & 10 & 50 & 100 & 392 & 13 \\
\hline
\texttt{gal4} & 3000 & 6 & 4 & 9764 & 12s & 0.67 & 7606 & 67s & 10 & 49 & 99 & 16 & 16\\
\hline
\texttt{gal4}$^{(*)}$ & 3000 & 6 & 4 & 9764 & 12s & 0.65 & 11733 & 191s & 10 & 50 & 100 &  9764 & 301\\
\hline
\texttt{uasgaba} & 1000 & 2 & 2 & 4586 & 30s & 0.70 & 4194 & 90s & 10 & 50 & 100 & 175 & 175\\
\hline
\end{tabular}
\end{center}
\caption{Results of the comparison with \varun.}\label{tab:madmx&varun}
\end{table}

For each experiment, we compared the best top-$m$ z-scores, with
$m=10, 50$, and $100$, as follows. Note that, in general, the top-$m$
motifs found by \madmx\ and \varun\ differ. Thus, we let $z_{M}^{i}$
(resp., $z_{V}^{i}$) be the z-score of the $i$th motif in 
decreasing z-score order obtained
by \madmx\ (resp., \varun). For each $m$, the table reports how many times it
was $z_{M}^{i} \geq z_{V}^{i}$, for $1 \leq i \leq m$. Also, column
$m^*$ (resp., column $\hat{m}$) gives the maximum $m$ such that
$z_{M}^{i} \geq z_{V}^{i}$ (resp., $z_{M}^{i} > z_{V}^{i}$) for every
$1 \leq i \leq m$.

Even when \madmx\ is calibrated to yield a slightly smaller output,
the quality of the motifs extracted, as measured by the z-score, is
higher than those output by \varun. Indeed, for sequences
\texttt{ace2} and \texttt{uasgaba} a very large prefix of the
top-ranked motifs extracted by \madmx\ features strictly greater
z-scores of the corresponding top-ranked ones extracted by \varun. In
fact, for all of the four sequences, at least the thirteen top-ranked
motifs enjoy this property.  To shed light on the slightly worse
performance of \madmx\ on \texttt{gal4}, we re-ran \madmx\ with a
different density threshold, so to obtain a slightly larger output
(see row \texttt{gal4}$^{(*)}$). In this case, the top-$301$ motifs
extracted by \madmx\ have z-score strictly greater than the
corresponding motifs extracted by \varun, while the execution time
remains still acceptable.

For all runs, the top z-score of a motif discovered by \madmx\ is
considerably higher than the one returned by \varun. Specifically, on
\texttt{ace2} our best z-score is 387\,763 vs.\mbox{} 12\,027 of
\varun; on \texttt{ap1}, we have 12\,027 vs.\mbox{} 1\,490; on
\texttt{gal4} it is 75 vs.\mbox{} 28; on \texttt{gal4}$^{(*)}$ it is
150 vs.\mbox{} 28; on \texttt{uasgaba} we have 134\,532 vs.\mbox{}
67\,059. This reflects the high selectivity of \madmx, which is to be
attributed mostly to adoption of a more flexible density constraint.

We must remark that \madmx\ (in its current nonoptimized
version) is slower than \varun, but it still runs in time
acceptable from the point of view of a user. To further investigate the
tradeoff between execution time and significance of the discovered
motifs, we repeated the experiments running \madmx\ with $\min_\ell=2$
and $\rho=0.65$, for all sequences. The running time of \madmx\ was
almost halved, while the small output produced still featured high
quality. In fact, for sequences \texttt{ace2}, \texttt{ap1}, and \texttt{uasgaba} the top-$100$ motifs extracted by \madmx\ have z-score greater or equal than the corresponding ones returned by \varun.

We also have attempted a comparison between \varun\  and \madmx\ 
on longer sequences (such as \HGO) at higher frequencies
(since, unfortunately, \varun\ does not seem to be able to handle low
frequencies on very long sequences). Even allowing a higher number of don't cares between solid characters ($D=2$) for the
motifs of \varun, 
all of the top-$m$ z-scores featured by the motifs extracted by \madmx\
are greater than or equal to the
corresponding scores in the ranking of \varun, with $m$ reaching the size of
\varun's output. 
In fairness, we remark that \varun\ was designed to work at its best
on protein sequences, while \madmx's main target are
\dna\ sequences. Hence, these two tools should be regarded as
complementary. Moreover, \varun\ has the advantage of retrieving
flexible motifs, while \madmx\ focuses only on rigid ones.

\paragraph{\bf Acknowledgments} 
The authors wish to thank Alberto Apostolico and Matteo Comin for providing
the code and giving valuable insights on \varun, Ben Raphael for
suggesting the use of \rmasker, and Roberta Mazzucco and Francesco
Peruch for coding \madmx.

\bibliographystyle{plain}

\end{document}